\documentclass[amsmath,amssymb,tightenlines,showpacs,floatfix]{revtex4}
\usepackage{graphicx}
\begin{document}
\title{Aharonov-Bohm differential conductance modulation in defective metallic single-walled carbon nanotubes}
\author{Mehran Bagheri}
\email{mh-bagheri@cc.sbu.ac.ir.}%
\affiliation{Physics Department, Shahid Beheshti University, Evin,
Tehran 19839, Iran}
\date{\today}
\begin{abstract}
Using a perturbative approach, the effects of the energy gap induced
by the Aharonov-Bohm (AB) flux on the transport properties of
defective metallic single-walled carbon nanotubes (MSWCNTs) are
investigated. The electronic waves scattered back and forth by a
pair of impurities give rise to Fabry-Perot oscillations which
constitutes a coherent backscattering interference pattern (CBSIP).
It is shown that, the CBSIP is aperiodically modulated by applying a
magnetic field parallel to the nanotube axis. In fact, the AB-flux
brings this CBSIP under control by an additional phase shift. As a
consequence, the extrema as well as zeros of the CBSIP are located
at the  irrational fractions of the quantity
$\Phi_\rho={\Phi}/{\Phi_0}$, where $\Phi$ is the flux piercing the
nanotube cross section and $\Phi_{0}=h/e$ is the magnetic quantum
flux. Indeed, the spacing between two adjacent extrema in the
magneto-differential conductance (MDC) profile is decreased with
increasing the magnetic field. The faster and higher and slower and
shorter variations is then obtained by metallic zigzag and armchair
nanotubes, respectively. Such results propose that defective
metallic nanotubes could be used as magneto-conductance switching
devices based on the AB effect.
\end{abstract}
 \maketitle
\section{Introduction}
\label{sec1}\indent Due to their quasi-one-dimensional structure and
intriguing electronic properties, carbon nanotubes have been
attracted an increasing amount of attentions\cite{R1}. Carbon
nanotubes are tubular nano-objects which can be thought of as
graphenes wrapped onto a seamless cylinder. Depending sensitively on
the wrapping vector, a nanotube may be either a one-dimensional (1D)
metal with a finite density of states at the Fermi energy or a
semiconductor with a gap. In special, for the sake of the 1D nature
of their electronic conduction bands near the Fermi energy, metallic
single-walled nanotubes constitute a nearly perfect realization of
1D quantum wires\cite{R2,R3,R4,R5}.\\
\indent The investigation of quantum transport in carbon nanotubes
is expected to have unprecedented potential applications for
developing nanoelectronic devices. They can be applied as conducting
quantum wires\cite{R5,R6}, single-electron tunneling
tranistors\cite{R7,R8}, field-effect transistors\cite{R9}, and
spin-electronic devices\cite{R10}. Theoretical calculations based on
the Landauer-B\"{u}tticker formalism\cite{R11,R12} predict the
conductance quantization for a perfect metallic nanotube for the
case of ideal contacts. The maximum value of the conductance near
the Fermi energy reaches $2G_0$, where $G_0=2e^{2}/h$ is the
conductance quantum\cite{R13}. However, in contrast to the pristine
nanotube, several theoretical
works\cite{R14,R15,R16,R17,R18,R19,R20} and experimental
evidences\cite{R21,R22,R23,R24} have shown that in the presence of
disorders coming from various sources like chemical impurities,
topological defects, Stone and Wales\cite{R25}, and vacancies this
quantized conductance of the nanotube does not follow the aforesaid
results. Practically, these imperfections are unavoidable in
manipulation nanotubes into devices and induce departure from
ballistic transport, and yet preserve quantum interference effects,
which can be profoundly affected by magnetic fields.\\
\indent Owing to the decoherence, the quantum corrections to the
classical conductance of a device are usually negligible in
macroscopic systems at the room temperature. In mesoscopic systems
at low temperatures, however, the quantum mechanical coherency
becomes more important because the phase coherence length $l_\phi$
increases with decreasing temperature. When the coherence length
$l_\phi$ exceeds the elastic mean free path $l_{m}$, scattering on
different impurities can interfere. Several quantum interference
(QI) modifications are (1) the WL correction, which originates from
pairs of time-reversed paths in a diffusive sample interfere
constructively in the zero magnetic field. This interference
enhances (reduces) the probability of electronic backscattering,
decreasing (increasing) the conductance of the sample\cite{R26}; (2)
the AB and Altshuler-Aronov-Spivak (AAS) oscillations. The AAS
effect is actually the same WL correction embracing a magnetic
field\cite{R27,R28,R29}. As the magnetic field is increased, the AB
phase eliminates the WL constructive interference, leading to a
magneto-conductance; (3) universal conductance fluctuations (UCF),
which means that the conductance fluctuations are independent of the
conductor details.\\
\indent Furthermore, one of the unique properties of carbon
nanotubes is that their metallicity can be controlled by an external
magnetic field applied parallel to the nanotube axis. This magnetic
field gives rise to a periodic energy gap at the charge neutrality
point (CNP), where the bonding and antibonding bands are crossed.
When the cross section of a nanotube is pierced by the magnetic
field, the electronic wavefunctions acquire an additional phase
$2\pi{\Phi}/{\Phi_0}$. Thus, metallic nanotubes can be made
semiconducting and vice versa. Over the past few years, remarkable
efforts have been undertaken to evidence the effects of a magnetic
field on the band structure of
nanotubes\cite{R30,R31,R32,R33,R34,R35,R36,R37,R38,R39,R40,R41,R42,R43}.\\
\indent Following our previous paper\cite{R44}, in which a
perturbative approach is well developed to include effects of the
band structure and impurity on transport characteristics of metallic
nanotubes, the current work concentrates on elucidating influences
of the AB-flux\cite{R45} on the differential conductance (DC). The
motivation of this attempt is to study how the magnetic field
dependence of the band-structure of the nanotube influences the DC.
This may provide us the possibility of fabricating
magneto-conductance switching devices based on the AB effect in
defective metallic nanotubes. It is shown that, for a couple of
impurities the nanotube behaves like a Fabry-Perot electron
resonator\cite{R53,R54} and the CBSIP resulting from the Fabry-Perot
oscillations is aperiodically modulated in the presence of the
AB-flux. Aperiodicity means that no specific magnetic flux
periodicity is found in the MDC profile. Further, extrema as well as
zeros of the MDC are positioned at irrational fractions of the
magnetic flux with
a spacing which is decreased by increasing the magnetic field.\\
\indent The paper is established as follow. In Sec. \ref{sec2}, the
model of Ref.\cite{R44} is developed to include the AB-flux. In Sec.
\ref{sec3}, we discuss the CBSIP in the presence of the AB-flux both
for a single and for a couple of impurities.
\section{THEORETICAL MODEL}
\label{sec2} 
\begin{figure}[htb]
\includegraphics[width=15cm,height=8cm,angle=0]{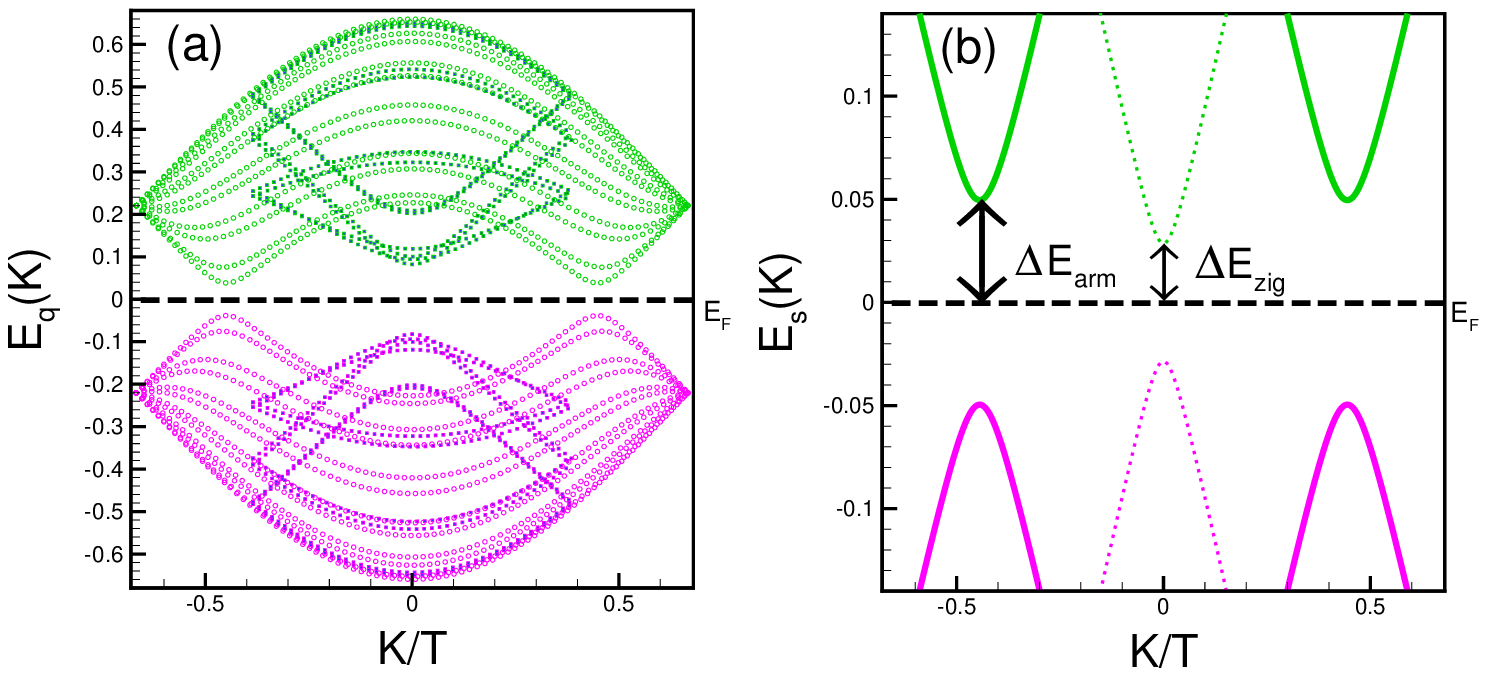}
{\caption{(Color online) The energy dispersion relation for subbands
of the armchair and metallic zigzag nanotubes in the presence of an
axial magnetic field. (a) A multiband model, coming from evaluating
Eq. (\ref{eq3}) in the presence of a $1000$ Tesla magnetic field
pointing along its axis, for the $(6,6)$ armchair (circle) and
$(6,0)$ zigzag (square) nanotubes. The nanotube now has a finite
subband-gap $\Delta_{g}(\Phi_{\rho})$ expressed by Eq. (\ref{eq4}),
and that all degenerate levels have been split. Antibonding bands
(green-$E_{q}(k)>0$) are symmetric to the bonding bands
(purple-$E_{q}(k)<0$). (b) A two-band model, which comes from
evaluating Eq. (\ref{eq5}) ($E_{s}(k)$ with $s=1$) in the presence
of a $10$ milli-Tesla magnetic field pointing along its axis,
includes the $(6,6)$ armchair (solid line) and $(6,0)$ zigzag
(dotted) nanotubes. The subband-gap is now expressed by
$\Delta_{g}(\Phi_{\rho})=3\Delta_{0}\Phi_{\rho}$. The electron
scattering processes change electrons from the right moving to the
left moving leading to electrical resistance. Generally, both
intrasubband and intersubband scattering events are likely. Energies
are scaled in
Rydberg and lengths in Bohr radius.}%
\label{fig1}}
\end{figure}

We address a defective MSWCNT in the presence of an
axial electric and magnetic field. The full descriptions of the
model in the absence of the magnetic field can be found in Ref.
\cite{R44}.  Here, we just add the AB-flux in its band-structure, so
the Hamiltonian of the whole system is given by
\begin{equation}
\label{eq1} \hat{{\cal {H}}}(\Phi_{\rho})=\hat{{\cal
{H}}}_{tube}(\Phi_{\rho})+\hat{{\cal {H}}}_{sd}+\hat{{\cal
{H}}}_{imp}.
\end{equation}
In the above equation the first term, describing the kinetic energy
of electrons for a perfect nanotube, is given by\cite{R44,R46}
\begin{equation}
\label{eq2} \hat{{\cal {H}}}_{tube}(\Phi_{\rho})=
\sum_{\alpha=\pm}\sum_{q=1}^{N_{t}/2}\sum_{k\in}^{FBZ}{\cal{E}}^{\alpha}_{q+\Phi_{\rho}}(k)\hat{C}^{\dag
\alpha}_{q}(k)\hat{C}^{\alpha}_{q}(k).
\end{equation}
In the presence of an uniform magnetic field $\vec B$ parallel to
the nanotube axis, the wrapping modes are modified according to
${q}/{r_t}\rightarrow{q}/{r_t}+{\Phi_{\rho}}/{r_t}$\cite{R31}, so
the magnetic field dependent band-structure
${\cal{E}}^{\pm}_{q+\Phi_{\rho}}(k)$ is\cite{R44}
\begin{eqnarray}
\label{eq3}%
\frac{{\cal{E}}^{\pm}_{q+\Phi_{\rho}}(k)}{\gamma_{0}}=&&\pm\left\{1+
4\cos\left[\frac{\sqrt{3}}{2}a_{cc}\left(\frac{1}{r_t}[q+\Phi_{\rho}]\sin\omega+
k\cos\omega\right)\right] \right.
\cos\left[\frac{3}{2}a_{cc}\left(\frac{1}{r_t}[q+\Phi_{\rho}]\cos\omega-
k\sin\omega\right)\right] \nonumber\cr \\
&& \left.
+4\cos^{2}\left[\frac{\sqrt{3}}{2}a_{cc}\left(\frac{1}{r_t}[q+\Phi_{\rho}]\sin\omega+
k\cos\omega\right)\right]\right\}^{\frac{1}{2}}\hspace{-3mm},
\end{eqnarray}
where operators $\hat{C}^{\dag\pm}_{q}(k)$ and
$\hat{C}^{\pm}_{q}(k)$ create and destroy electrons in the orbital
with energy ${\cal{E}}^{\pm}_{q+\Phi_{\rho}}(k)$, respectively. The
$+$ and $-$ signs correspond to the conduction and valence band,
respectively. Good quantum numbers of electron states are $(q,k)$
where $q=1,\ldots,{N_{t}/2}$ and
$k\in\left(-{\pi}/{T},{\pi}/{T}\right)$. The quantities $N_t$,
$\mathcal{N}={N_t}/{2}$, $T$, $r_{t}$, $a_{cc}\simeq 1.44${\AA}, and
$\gamma_{0}\simeq 3.0$ eV are the number of carbon atoms in the
nanotube unit cell, the number of graphene unit cells in a given
nanotube unit cell, the length of the translation vector, the
nanotube radius, the C-C bond length, and the nearest-neighbor
overlap integral energy, respectively. Also,
$\omega={\pi}/{6}-\theta$ where $\theta$ is the chiral angel of the
nanotube whose value for the armchair and zigzag nanotube is
${\pi}/{6}$ and $0$, respectively. It is assumed that the on-site
energy is zero and the Fermi energy remains unchanged at the CNP. In
the zero magnetic field, all metallic linear bands cross the undoped
Fermi level either degenerated at $k_F=0$ (metallic zigzag) or
separated at $k_F=\pm{2\pi}/{3T}$ (armchair) in the first Brillouin
zone (FBZ). $\Phi_{\rho}$ equals ${\Phi}/{\Phi_0}$, with $\Phi=\pi
r_{t}^2B$. When $\Phi_{\rho}$ becomes an integer, the AB-flux is
canceled by $q$. It means that the gap induced by the magnetic field
oscillates periodically and can be obtained by the expression
$\Delta_{g}(\Phi_{\rho})=2\min\{\mid{\cal{E}}^{+}_{q+\Phi_{\rho}}(k)\mid\}$(see
Fig.~\ref{fig1}a). Lu\cite{R42} has shown that, for metallic
nanotubes the energy gap induced by an axial magnetic field is
expressed by
\begin{eqnarray}
\label{eq4}%
\Delta_{g}(\Phi_{\rho})= \begin{cases}
3\Delta_{0}\Phi_{\rho},&if\hspace{3mm}0\leq\Phi_{\rho}\leq\frac{1}{2}\\ \\
3\Delta_{0}|1-\Phi_{\rho}|,&if\hspace{3mm}\frac{1}{2}\leq\Phi_{\rho}\leq1,
\end{cases}
\end{eqnarray}
where $\Delta_{0}={\gamma_{0}a_{cc}}/{r_t}$ defines a characteristic
energy associated with the nanotube. Note that the expression
${\cal{E}}^{\pm}_{q+\Phi_{\rho}}(k)=\pm\gamma_{0}\sin[{\pi(q+\Phi_{\rho})}/n]$
gives van Hove singularities (VHSs) positions. Further, for later
calculations, we have exploited the corresponding Bloch's states of
an isolated nanotube previously derived in Ref.\cite{R44}.\\
\indent For considering the magneto-transport properties of the
nanotube near the Fermi level, we adopt the light-cone approximation
of the dispersion relation of Eq. (\ref{eq3}) which provides us a
simple formula of the $s$-th 1D subband around $k_F$. Thus, Eq.
(\ref{eq3}) reduces to\cite{R31}
\begin{eqnarray}
\label{eq5}%
\frac{{\cal{E}}^{\pm}_{s+\Phi_{\rho}}(k)}{\gamma_{0}}=\pm\frac{3a_{cc}}{2}\left[\left(\frac{s-1}{r_t}+\frac{\Phi_{\rho}}{r_t}\right)^2
+(k\mp k_F)^2\right]^{{1}/{2}},
\end{eqnarray}
where $s=1,\ldots,N_{t}/2$. For the lowest lying subband, with $s=1$
around $k=\pm k_F$, the energy band gap in the absence of the
magnetic field is zero. Using eq. (\ref{eq5}) one obtains
$\Delta_{g}(\Phi_{\rho})=3\Delta_{0}\Phi_{\rho}$. As the field
strength increases the line through the Fermi energy at zero
magnetic field is shifted away further from the CNPs thus given rise
to an increasing energy-gap. It is also worth mentioning that, the
quantity $\mu_{orb}=ev_{F}r_{t}/2$, with
$v_{F}={3\gamma_{0}a_{cc}}/{2\hbar}\approx 10^{6} m/s$, is the
magnetic moment of an electron traveling in a loop of radius $r_t$
with velocity $v_{F}$. Changes in the energy of electron states can
be described by the interaction of this orbital magnetic moment with
an axial magnetic field. A magnetic field parallel to the nanotube
axis is predicted to shift the energy of these states by $\Delta
E=-{\vec \mu_{orb}\cdot\vec B}=\pm
ev_{F}r_{t}B/2=\pm{3\Delta_{0}\Phi_{\rho}}/{2}$ (see
Fig.~\ref{fig1}(b)).\\
\indent Furthermore, the second and third terms in equation (1) are,
respectively, the Hamiltonian of non-interacting electrons under the
external source–drain voltage $V_{sd}$ and the Hamiltonian of the
interaction of electrons with impurities \cite{R47,R48,R49} like
those presented in \cite{R44}. Eventually, upon substituting $q$ in
Eq. (20) of Ref.\cite{R44} by $q+\Phi_{\rho}$, we obtain the
dimensionless form of the MDC at the zero temperature as follow
\begin{eqnarray}
\label{eq6}
\frac{G_{imp}^{\alpha\alpha}[V_{sd},{\cal{E}_{F}}(0),\Phi_{\rho}]}{G_0}&=&\frac{\pi^2
}{2}\sum_{q,q'=1}^{N_{t}/2}\sum_{k,k'\in}^{
FBZ}\sum^{r}_{\xi,\eta=1}J^{q
q'}_{\xi,\alpha\alpha}(k,k')J^{q'q}_{\eta,\alpha\alpha}(k',k)\nonumber\cr \\
&&\times\delta\Big[{\cal{E}}^{\alpha}_{q+\Phi_{\rho}}(k)-{\cal{E}}^{\alpha}_{q'+\Phi_{\rho}}(k')\Big]\nonumber\cr\\
&&\times\Big[sign[v^{\alpha}_{q+\Phi_{\rho}}(k)]sign[v^{\alpha}_{q'+\Phi_{\rho}}(k')]-1\Big]\nonumber\cr\\
&&\times\left\{\delta\Big[{\cal{E}_{F}}(0)-{\cal{E}}^{\alpha}_{q+\Phi_{\rho}}(k)-\frac{eV_{sd}}{2}sign[v^{\alpha}_{q+\Phi_{\rho}}(k)]\Big]\right.\nonumber\cr \\
&&+\left.\delta\Big[{\cal{E}_{F}}(0)-{\cal{E}}^{\alpha}_{q'+\Phi_{\rho}}(k')-\frac{eV_{sd}}{2}sign[v^{\alpha}_{q'+\Phi_{\rho}}(k')]\Big]\right\},
\end{eqnarray}
where $G^{\alpha\alpha}_{total,imp}=G^{++}_{imp}+G^{--}_{imp}$.
Also, $v^{\pm}_{q+\Phi_{\rho}}(k)=({1}/{\hbar}){\partial {{\cal
{E}}}^{\pm}_{q+\Phi_{\rho}}(k)}/{\partial k}$ is the electron
velocity, and $J^{q q'}_{\xi,\alpha\beta}(k,k')$ is a matrix for the
impurity potential located at a position, namely,
$\vec{x}_{\xi}$\cite{R44}. We have also assumed that the magnetic
field does not affect the Fermi energy, i.e.
${\cal{E}_{F}}(B)={\cal{E}_{F}}(0)$. More importantly, the
expression
$[sign[v^{\alpha}_{q+\Phi_{\rho}}(k)]sign[v^{\alpha}_{q'+\Phi_{\rho}}(k')]-1]$
controls the scattering event from the initial state to the final
state via the sign of the electron velocity. It requires that only
backward scattering events are possible in one-dimensional systems
like nanotubes. The coherent backscattering (CBS) of the electron is
an effect that describes the appearance of a backscattered peak when
the electron traveling in a time-reversed path self-interferes
constructively in the backscattered direction. It means that the
electronic wave is weakly localized\cite{R50,R51,R52}.\\
\indent By obtaining the solutions of the energy-momentum
conservation equation, i.e.
${\cal{E}}^{\alpha}_{q+\Phi_{\rho}}(k)={\cal{E}}^{\alpha}_{q'+\Phi_{\rho}}(k+\mathbf{g})$
where $\mathbf{g}$ is the transferred momentum, we now evaluate Eq.
(\ref{eq6}) at some special k-points in the FBZ. Using Eq.
(\ref{eq3}) for the $(n,n)$ armchair nanotubes one obtains
\begin{eqnarray}
\label{eq7}\mathbf{g}^{\pm}&&=-k\pm\frac{2}{\sqrt{3}a_{cc}}\arccos\left\{-\frac{1}{2}\cos\left(\frac{3(q'+\Phi_{\rho})a_{cc}}{2r_t}\right)\right.\nonumber\cr \\
&& \left.
\pm\frac{1}{2}\sqrt{\cos^{2}\left(\frac{3(q'+\Phi_{\rho})a_{cc}}{2r_t}\right)
+4\cos^{2}\left(\frac{\sqrt{3}ka_{cc}}{2}\right)+4\cos\left(\frac{\sqrt{3}ka_{cc}}{2}\right)
\cos\left(\frac{3(q+\Phi_{\rho})a_{cc}}{2r_t}\right)}\right\},
\end{eqnarray}
and for the $(n,0)$ zigzag nanotubes the equivalent expression is
given by
\begin{eqnarray}
\label{eq8}\mathbf{g}^{\pm}&&=-k\pm\frac{2}{3a_{cc}}\arccos\left\{\frac{1}{\cos\left(\frac{\sqrt{3}(q'+\Phi_{\rho})a_{cc}}{2r_t}\right)}\right.\nonumber\cr \\
&& \left.
\times\left[-\cos^{2}\left(\frac{\sqrt{3}(q'+\Phi_{\rho})a_{cc}}{2r_t}\right)+\cos^{2}\left(\frac{\sqrt{3}(q+\Phi_{\rho})a_{cc}}{2r_t}\right)+
\cos\left(\frac{\sqrt{3}(q+\Phi_{\rho})a_{cc}}{2r_t}\right)\cos\left(\frac{3ka_{cc}}{2}\right)\right]\right\}.
\end{eqnarray}
For the intrasubband scattering, i.e. $|q+\Phi_{\rho},k\rangle
\rightarrow
|q'+\Phi_{\rho},k'\rangle=|q+\Phi_{\rho},k+\mathbf{g}\rangle$, Eq.
(\ref{eq7}) has four scattering roots as follow
\begin{eqnarray}
\label{eq9}\mathbf{g}^{\pm}&&=0,\nonumber \\
&&-2k,\nonumber\\
&&-k\pm\frac{2}{\sqrt{3}a_{cc}}\arccos\left[\cos\left(\frac{3(q+\Phi_{\rho})a_{cc}}{2r_t}\right)+
\cos\left(\frac{\sqrt{3}ka_{cc}}{2}\right)\right],\nonumber \\
\end{eqnarray}
while for the metallic zigzag nanotubes Eq. (\ref{eq8}) provides
only two roots $0$ and $-2k$.\\
\underline{$k'=k$}: The root $\mathbf{g}^{\pm}=0$ means that $q$ and
$k$ are conserved, and no scattering event is occurred. Thus, the MDC becomes zero.\\
\underline{$k'=-k$}: The root $\mathbf{g}^{\pm}=-2k$ describes the
CBS of the electron within the same subband to another Fermi point.
In the CBS effect, the electron is elastically scattered back to a
momentum directly opposite to its original momentum state in the
momentum space. Let later on replace $G_{imp}$ by $G_{CBS}$. For a
${\it couple}$ of impurities located at $\vec
x_{\xi}=\vec{T}_{l_{1}}+\vec{R}_{j_{1}}+\vec{d}_1$ and $\vec
x_{\eta}=\vec{T}_{l_{2}}+\vec{R}_{j_{2}}+\vec{d}_2$, Eq. (\ref{eq6})
yields\cite{R44}
\begin{eqnarray}
\label{eq10}%
\Re\left(\frac{G_{CBS}^{\alpha\alpha}[V_{sd},{\cal{E}_{F}}(0),\Phi_{\rho}]}{G_0}\right)&=&e|V_{sd}|\Big(\frac{\pi
g }{2\mathcal{MN}}\Big)^{2}\sum_{q=1}^{N_{t}/2}\sum_{k\in}^{FBZ}
\delta\left\{\Big[{\cal{E}_{F}}(0)-{\cal{E}}^{\alpha}_{q+\Phi_{\rho}}(k)\Big]^{2}-\left[\frac{eV_{sd}}{2}\right]^{2}\right\}\nonumber\\
&&\times\cos\left\{2k\left[(l_{2}-l_{1})T+\left(\vec{R}_{j_{2}}-\vec{R}_{j_{1}}\right)\cdot\frac{\vec{T}}{T}\right]\right\},
\end{eqnarray}
where $\mathcal{M}$ is the total number of nanotube unit
cells\cite{R44}. Because
${\cal{E}}^{+}_{q+\Phi_{\rho}}(k)=-{\cal{E}}^{-}_{q+\Phi_{\rho}}(k)$;
if ${\cal{E}}_{F}(0)=0$ then ${G_{CBS}^{++}}={G_{CBS}^{--}}$. For
the case of a {\it single} impurity the CBSIP is killed. Because two
carbon atoms $A$ and $B$ inside a graphite unit cell belong to two
different sublattices, the impurity can occupy one of the lattice
site. For simplicity, we have here assumed that two impurities are
substituted on $B-$sites with the same circumferential angle along
the nanotube axis\cite{R44}. These arrangements of impurities break
all mirror symmetry planes containing the nanotube axis\cite{R16}.
By turning the sum over k into an integral and with exploiting Eq.
(\ref{eq5}) for the lowest lying subband, Eq. (\ref{eq10}) leads to
\begin{eqnarray}
\label{eq11}
\Re\left(\frac{G_{CBS}^{\alpha\alpha}[V_{sd},\Phi_{\rho}]}{G_0}\right)&=&\left(\frac{\pi
eV_{sd}
g^{2}T^{Y}}{X\hbar^2v_{F}^{2}\mathcal{M}{\mathcal{N}}_{Y}^{2}}\right)\left[\left(\frac{eV_{sd}}{\hbar
v_F}\right)^{2}-\left(\frac{\Phi_{\rho}^{Y}}{r_{t}^{Y}}\right)^{2}\right]^{-\frac{1}{2}}\nonumber\\
&&\times\cos\Big[2k_{F}(l_2-l_1)T^{Y}\Big]
\cos\left[\sqrt{\left(\frac{eV_{sd}}{\hbar
v_F}\right)^{2}-\left(\frac{\Phi_{\rho}^{Y}}{r_{t}^{Y}}\right)^{2}}(l_{2}-l_{1})T^{Y}\right].
\end{eqnarray}
The total DC is then $\Re[{G_{CBS}^{tot,Y}}]=
2\Re[{G_{CBS}^{++,Y}}]=2\Re[{G_{CBS}^{--,Y}}]$. From Eq.
(\ref{eq11}) one can draw several conclusions: (1) for the armchair
nanotubes we have $X=1$, $Y=arm$, and $k_{F}={2\pi}/{3T^{arm}}$,
while for the metallic zigzag ones $X=2$, $Y=zig$, and $k_{F}=0$;
(2) the cosine term is responsible for the CBSIP. Averaging over
different impurity configurations melts away this interference term;
(3) no switching effect from positive to negative MDC is occurred by
changing the orientation of the magnetic field with respect to the
nanotube axis. This means that, the reciprocity relation
$G_{CBS}(\Phi_{\rho})=G_{CBS}(-\Phi_{\rho})$ is fulfilled; (4) the
amplitude of this CBSIP depends on both the source-drain voltage and
the AB-flux; (5) in the limit $\Phi_{\rho}\rightarrow 0$, one
recovers the solution of the free-magnetic field case derived in
Ref.\cite{R44}; (6) conduction through this gapped nanotube is
dependent sensitively on the exact position of the $V_{sd}$ with
respect to the lowest level subband edges. Strictly speaking, there
is a threshold voltage determined by
$eV_{sd}\geq3\Delta_{0}\Phi_{\rho}/2$ and
$eV_{sd}\leq-3\Delta_{0}\Phi_{\rho}/2$, below and above which,
respectively, the transport is forbidden. This issue is in agreement
with the density of state due to the one dimensional subbands
expected for semiconductor nanotubes. In other words, the MDC is
singular at the position of the lowest subband bottom indicating its
van-Hove singularity; (7) a closer look at the argument of the
second cosine term reveals that the interference term leads to
aperiodic oscillations in the MDC profile. This is because this
argument is a nonlinear mapping of the AB-flux as well as the
source-drain voltage. In fact, the DC is aperiodically modulated
through the AB-flux. At zero temperature, it would be plausible if
we suppose that the system size plays the role of the
phase-coherence length. In the presence of the AB-flux the electrons
acquire additional phases, and we can control the interference
pattern made from the conjugated time-reversed paths. More
importantly is the negative MDC. Actually, it originates from not
only the QI effects but also the pseudospin conservation rule. The
negative MDC feature may be exploited for designing
magneto-conductance switches based on the AB effect.\\
\underline{$k'=\pm({2}/{\sqrt{3}a_{cc}})\arccos\left[\cos({3(q+\Phi_\rho)a_{cc}}/{2r_t})+
\cos({\sqrt{3}ka_{cc}}/{2})\right]$}: These two last roots are
actually the intersubband backscattering around the the same Fermi
point, and we currently discard them\cite{R56}.
\section{DISCUSSIONS}
\label{sec3} 
\begin{figure}[htb]
\includegraphics[width=15cm,height=8cm,angle=0]{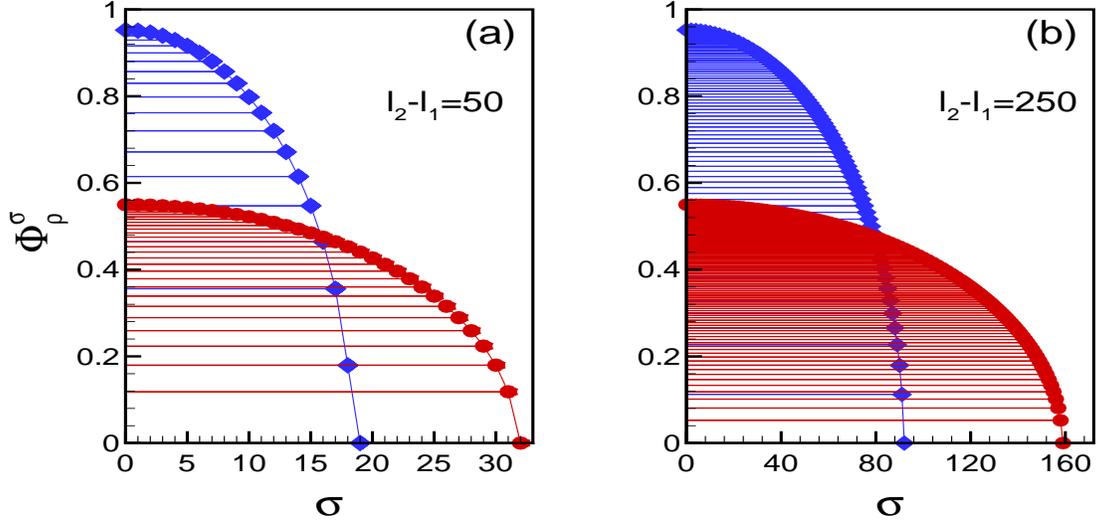}
{\caption{(Color online) The positions of extrema in the
$\Phi_{\rho}$-space are calculated with using the upper part of Eq.
(\ref{eq14}). (a) The allowed $\sigma$'s for the $(6,6)$
(diamond-blue) and $(6,0)$ (circle-red) nanotubes, with $l_2-l_1=50$
and $eV_{sd}=0.11$, are 18 and 33, respectively. (b) The allowed
$\sigma$'s for the $(6,6)$ (diamond-blue) and $(6,0)$ (circle-red)
nanotubes, with $l_2-l_1=250$ and $eV_{sd}=0.11$, are 93 and 160,
respectively. The spacing between two horizontally adjacent lines is
decreased with increasing the magnetic field which obviously shows
that oscillations are aperiodic.} \label{fig2}}
\end{figure}

\begin{figure}[htb]
\includegraphics[width=15cm,height=8cm,angle=0]{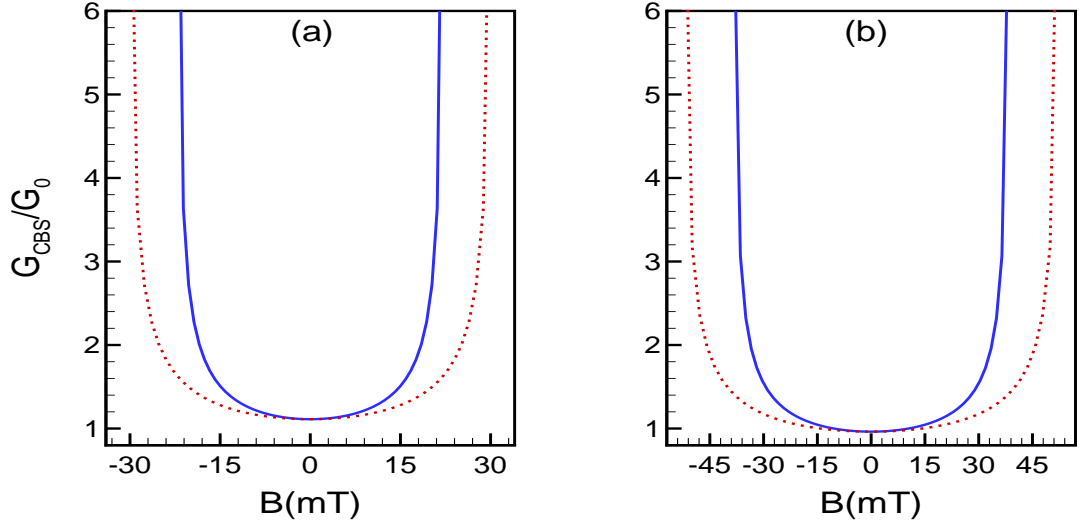}
{\caption{(Color online) Calculated MDC as a function of the
magnetic field $B$ for a single impurity. Results come from
evaluating Eq. (\ref{eq11}) for the armchair and metallic zigzag
nanotubes. (a) Traces are plotted for the $(6,6)$ armchair, with
$eV_{sd}=0.11$ and $B\in[-22,22]$ milli-Tesla (solid-blue), and
$eV_{sd}=0.15$ with $B\in[-30,30]$ milli-Tesla (dotted-red). (b)
Traces are plotted for the $(6,0)$ zigzag, with $eV_{sd}=0.11$ and
$B\in[-38,38]$ milli-Tesla (solid-blue), and $eV_{sd}=0.15$ with
$B\in[-52,52]$ milli-Tesla (dotted-red). They exhibit a U-like
behavior. The plateau
of the zigzag nanotube is approximately twice the plateau of the armchair one.}%
\label{fig3}}
\end{figure}

\begin{figure}[htb]
\includegraphics[width=15cm,height=12cm,angle=0]{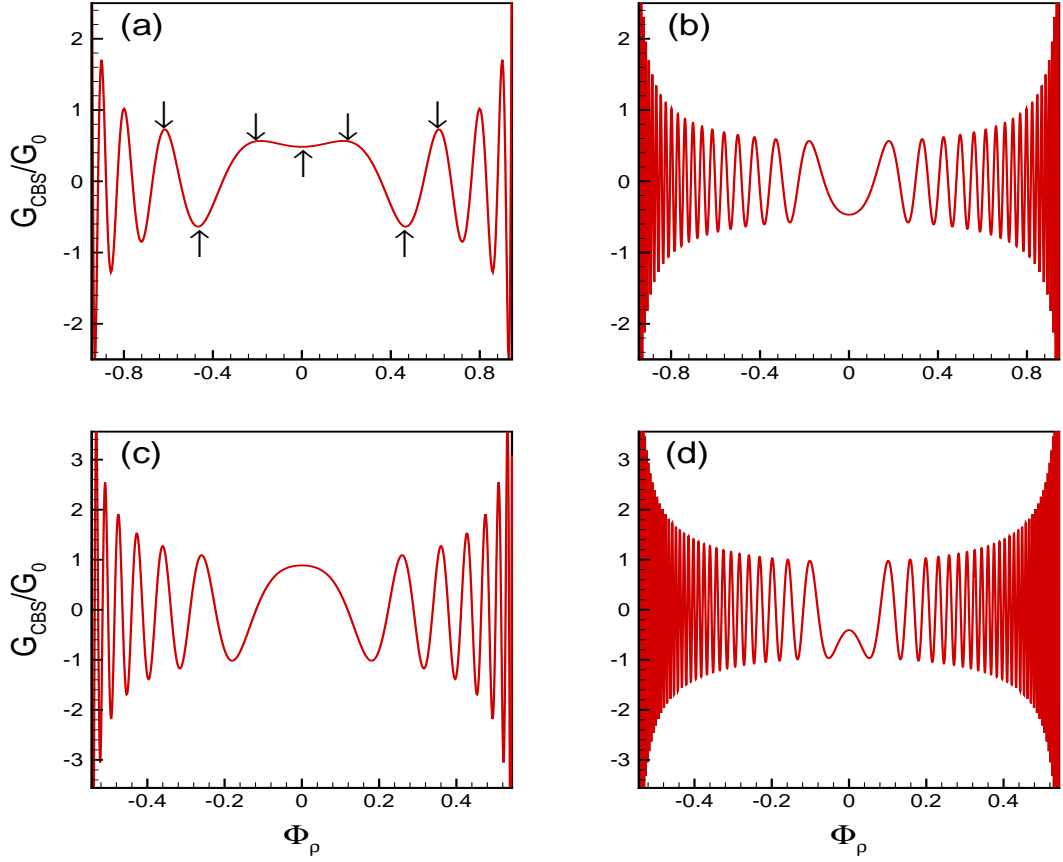}
{\caption{(Color online) The calculated MDC coming from evaluating
Eq. (\ref{eq11}) for the nanotube $(n,m)$ for a pair of impurities.
The CBSIP shows aperiodic oscillations. The extrema as well as zeros
are located at the irrational fractions of $\Phi_\rho$. Positions of
some extrema are indicated by arrows. (a) $(n,m)=(6,6)$,
$eV_{sd}=0.11$, $B\in[-22,22]$ (mT), and $l_2-l_1=50$; (b)
$(n,m)=(6,6)$, $eV_{sd}=0.11$, $B\in[-22,22]$ (mT), and
$l_2-l_1=250$; (c) $(n,m)=(6,0)$, $eV_{sd}=0.11$, $B\in[-38,38]$
(mT), and $l_2-l_1=50$; (d) $(n,m)=(6,0)$, $eV_{sd}=0.11$,
$B\in[-38,38]$ (mT), and $l_2-l_1=250$. A comparison between, say
panels (b) and (d), exhibits that the faster/higher and
slower/shorter
aperiodic fluctuations belong to metallic zigzag and armchair nanotubes, respectively.}%
\label{fig4}}
\end{figure}

Using the two-terminal Landauer-B\"{u}tticker approach
for a two-band model, the whole resistance of the nanotube is
approximately given by\cite{R12}
\begin{equation}
\label{eq12}
G_{tube}^{-1}=(G_{perfect})^{-1}+G_{CBS}^{-1}+G_{c1}^{-1}+G_{c2}^{-1}.
\end{equation}
In the above equation, the first term is the resistance of a perfect
ballistic nanotube with perfect contacts. It originates from the
redistribution of electrons between reservoirs and the nanotube. The
second terms is the quantum correction coming from the CBS effect.
Two last terms, discarded here, are for imperfect contacts between
the nanotube and reservoirs. To investigate the behavior of the MDC
as a function of the AB-flux, we have numerically performed Eq.
(\ref{eq11}) for both armchair and zigzag nanotubes. Results are the
same for both repulsive and attractive impurity potentials. Let us
suppose $g$=$10^{4}\gamma_{0}$ representing a typical impurity and
$\mathcal{M}=1000000$. In Eq. (\ref{eq11}), the product of two
cosine terms is actually a resultant wave coming from the
superposition of two standing waves with the same amplitude but
different wavenumbers $k_1=k_F+(1/2)\sqrt{({eV_{sd}}/{\hbar
v_F})^2-(\Phi_{\rho}/r_t)^2}$ and
$k_2=k_F-(1/2)\sqrt{({eV_{sd}}/{\hbar v_F})^2-(\Phi_{\rho}/r_t)^2}$.
These two initial standing waves describing two degenerate resonant
states induced by impurities in the $FBZ$ are given by
\begin{eqnarray}
\label{eq13} f_i=\left(\frac{\pi eV_{sd}
g^{2}T}{2X\hbar^2v_{F}^{2}\mathcal{M}{\mathcal{N}}^{2}}\right)\left[\left(\frac{eV_{sd}}{\hbar
v_F}\right)^{2}-\left(\frac{\Phi_{\rho}}{r_{t}}\right)^{2}\right]^{-\frac{1}{2}}\cos(2k_{i}l_m),\hspace{3mm}i=1,2.
\end{eqnarray}
Because two functions $f_1$ and $f_2$ are not periodic in the
$\Phi_\rho$-space, so is their superposition, i.e. $f_1+f_2$. Thus,
an aperiodic variation in the MDC is expected. The phase difference
for an electron propagating over the length $l_m$ is given by
$\delta\varphi(\Phi_{\rho})=2\Delta kl_{m}$ where $\Delta
k=k_1-k_2=\sqrt{({eV_{sd}}/{\hbar v_F})^2-(\Phi_{\rho}/r_t)^2}$.
Constructive interference occurs when the extrema of two waves add
together and the phase difference becomes an integer multiple of
$\pi$, i.e. $\delta\varphi(\Phi_{\rho})=\sigma\pi$, with
$\sigma\in\mathbb{Z}$. On the other hand, destructive interference
occurs when two waves have a phase difference of a half-integer
multiple of $\pi$, i.e.
$\delta\varphi(\Phi_{\rho})=(\sigma+1/2)\pi$. An analytic expression
in the $\Phi_{\rho}$-space can be derived easily as follow
\begin{eqnarray}
\label{eq14}%
\Phi_{\rho}^{\sigma}=\begin{cases} \pm
r_{t}\Big[\left(\frac{eV_{sd}}{\hbar
v_F}\right)^{2}-\left(\frac{\sigma\pi}{2l_m}\right)^{2}\Big]^{\frac{1}{2}},&\hspace{3mm}constructive\\ \\
\pm r_{t}\Big[\left(\frac{eV_{sd}}{\hbar
v_F}\right)^{2}-\left(\frac{(2\sigma+1)\pi}{2l_m}\right)^{2}\Big]^{\frac{1}{2}},&\hspace{3mm}destructive.
\end{cases}
\end{eqnarray}
The above equation gives actually the spacing between the MDC
extrema (constructive) or zeros (destructive)in the
$\Phi_{\rho}$-space. Due to the nonlinear mapping between
$\delta\varphi(\Phi_{\rho})$ and $\Phi_{\rho}$, the MDC vs.
$\Phi_{\rho}$ behaves aperiodically. The most important feature is
that, extrema and zeros are located at irrational fractions in the
$\Phi_{\rho}$-space. In other words, $\Phi=${\it
irrational}$\times\Phi_0$. The nonlinear dependence of the extrema
positions as a function of $\sigma$ is depicted in
Figs.~\ref{fig2}(a) and 2(b) for $l_2-l_1=50$ and 250, respectively.
In both panels the nonlinear behavior of oscillations can be seen by
comparing the spacing between two horizontally adjacent lines. It
should be pointed out that, variations in the MDC are aperiodic in
the $eV_{sd}$-space as well.\\
\indent Also, $\Delta k$ can be expressed in terms of the series
$\Delta
k=\lambda\left(1+{\chi}/{2}-{\chi^2}/{8}+{\chi^3}/{16}-\cdots\right)$,
where $\lambda=|{eV_{sd}}/{\hbar v_F}|$ and
$\chi(\Phi_{\rho})=-({\Phi_{\rho}}/{r_{t}})^2/({eV_{sd}}/{\hbar
v_F})^2$. Thus, the phase accumulated by the electron can be
expressed by
$\delta\varphi=\delta\varphi(0)+\delta\varphi(\Phi_{\rho})$, where
$\delta\varphi(0)=2\lambda l_m$ is the phase difference in the
absence of the AB-flux and $\delta\varphi(\Phi_{\rho})=\lambda
l_{m}\left({\chi}-{\chi^2}/{4}+{\chi^3}/{8}-\cdots\right)$ is the
magnetic field dependent phase difference. As a check, we see that,
for $\chi=0$, i.e. in the absence of the AB-flux,
$\delta\varphi(\Phi_{\rho})=0$. For the $(n,n)$ armchair and $(n,0)$
zigzag nanotubes we find $r_{t}^{arm}={3na_{cc}}/{2\pi}$;
$T^{arm}=\sqrt{3}a_{cc}$, and
$r_{t}^{zig}={\sqrt{3}na_{cc}}/{2\pi}$; $T^{zig}=3a_{cc}$,
respectively, so in case we approximate $\delta\varphi(\Phi_{\rho})$
by $\lambda l_{m}{\chi}$, it is straight to show that
$\delta\varphi(\Phi_{\rho})^{arm}/\delta\varphi(\Phi_{\rho})^{zig}=\sqrt{3}$.\\
\indent For a single impurity where $l_2=l_1$, the nanotube is less
defective. In this case the quantum interference due to the CBS is
killed, and the AB modulation becomes dominant. Evaluating Eq.
(\ref{eq11}) leads to a U-like behavior for the armchair and zigzag
nanotubes. As depicted in Fig.~\ref{fig3}, these curves are
symmetric and centered at $B=0$ (or $\Phi_{\rho}=0$). Each curve has
a plateau which decreases with increasing the magnetic field and its
magnitude depends strongly on the location of the source-drain
voltage. A deep look at upon panels 3(a) and 3(b) shows that the
magnitude of the zigzag plateau is approximately twice the armchair
one for a fixed value of the source-drain voltage. Recently,
Lassagne {\it et al} \cite{R40} have also observed such U-like
curves, of course through a Schottky barrier for different gate
voltages at non-zero temperatures, for a clean multi-walled nanotube
threaded by the AB-flux. We should emphasize that, although our
result share some similarities in the AB-pattern with that of
Ref.\cite{R40}, but their underlying physical transport phenomena
could be different. It is expected that, at a non-zero temperature
and gate voltage such U-like behavior would drastically changed in
our model.\\
\indent Moreover, for a couple of impurities, with $l_1\neq l_2$,
the MDC as a function of $\Phi_\rho$ for two different
interdistances between impurities is calculated. In Fig.~\ref{fig4},
panels show aperiodic fluctuations which alter between positive and
negative values. The amplitude of oscillations is increased with
increasing the magnetic field, while the spacing between two
adjacent extrema is decreased. These fluctuations represent a
hallmark of defective quantum transport resulting from the
competition between the CBS effect and the AB-flux. Such
fluctuations may be attributed to the Fabry-Perot
oscillations\cite{R53} modulated by the AB-flux. Positions of some
extrema in the MDC are labeled by arrows in Fig.~\ref{fig4}(a). The
most striking and immediately visible difference between armchair
oscillations and zigzag ones, say by comparing panels 4(b) and 4(d),
is that for the same value of the source-drain voltage the
fluctuations of the zigzag nanotubes are faster and higher than that
of the armchair one. The envelope functions of the extrema have a
U-form portrait as well.\\
\indent In summary, this semi-classical study shows the subtle
interplay between the quantum interference phenomena originating
from Fabry-Perot oscillations\cite{R53} and the magnetic
field-dependent of the band structure in defective metallic
nanotubes. We have shown that, how such oscillations can be
modulated using the AB-flux. Nonlinear mapping between the MDC and
the magnetic filed leads to aperiodic fluctuations. Such results may
be applied for manipulating defective metallic nanotubes into
quantum interference devices, say, for the construction of nanotube
magneto-conductance devices based on the AB effect\cite{R55}.
Moreover, it is worth mentioning that the model is flexible to
incorporate inelastic events like the electron-electron and
electron-phonon scattering events. In the presence of such
decoherent effects we expect a drastic change in the interference
pattern of the differential
conductance.\\
\section{Acknowledgements}
\label{sec4} The author would like to acknowledge Professor Stephan
Roche for reading the manuscript, and suggesting effective and
critical points..

\newpage
\newpage
\newpage
\newpage

\end{document}